\documentclass[letterpaper,twocolumn,10pt]{article}
\usepackage{usenix}
\usepackage{amsmath, soul}
\usepackage{multirow}
\usepackage{subfig}
\usepackage{pifont}
\usepackage{xcolor}
\usepackage{caption}
\usepackage{float}

\DeclareCaptionFormat{myformat}{#1#2#3\hrulefill}
\usepackage{comment}
\usepackage{graphicx}
\usepackage[normalem]{ulem}
\usepackage{url}
\newcommand{\cmark}{\ding{51}}%
\newcommand{\xmark}{\ding{55}}%
\graphicspath{ {./figures/} } 

\usepackage[utf8]{inputenc}
\usepackage{fancyhdr}
\usepackage{lastpage}

\definecolor{gatororange}{RGB}{250,70,22}
\definecolor{princess}{RGB}{255,123,156}
\definecolor{seafoam}{RGB}{78,255,172}

\newcommand{\captcha}{CAPTCHA}
\newcommand{\kvus}{Yeehaw Junction}
\newcommand{\kvog}{Kenansville}

\begin{document}
\title{Attacks as Defenses: Designing Robust Audio CAPTCHAs Using
Attacks on Automatic Speech Recognition Systems}

\author{
{\rm Hadi Abdullah}\\
University of Florida
\and
{\rm Aditya Karlekar}\\
University of Florida
\and
{\rm Saurabh Prasad}\\
University of Florida
\and
{\rm Muhammad Sajidur Rahman}\\
University of Florida
\and
{\rm Logan Blue}\\
University of Florida
\and
{\rm Luke A. Bauer}\\
University of Florida
\and
{\rm Vincent Bindschaedler}\\
University of Florida
\and
{\rm Patrick Traynor}\\
University of Florida
}

\maketitle
\begin{abstract}
    Audio CAPTCHAs are supposed to provide a strong defense for online
    resources; however, advances in speech-to-text mechanisms have
    rendered these defenses ineffective.  Audio CAPTCHAs cannot simply
    be abandoned, as they are specifically named by the W3C as important
    enablers of accessibility.  Accordingly, demonstrably more robust
    audio CAPTCHAs are important to the future of a secure and
    accessible Web.  We look to recent literature on attacks on
    speech-to-text systems for inspiration for the construction of
    robust, principle-driven audio defenses.  We begin by comparing 20
    recent attack papers, classifying and measuring their suitability to
    serve as the basis of new ``robust to transcription'' but ``easy for
    humans to understand'' CAPTCHAs. After showing that none of these
    attacks alone are sufficient, we propose a new mechanism that is
    both comparatively intelligible (evaluated through a user study) and
    hard to automatically transcribe (i.e., $P({\rm transcription}) = 4
    \times 10^{-5}$).  Finally, we demonstrate that our audio samples
    have a high probability of being detected as CAPTCHAs when given to
    speech-to-text systems ($P({\rm evasion}) = 1.77 \times 10^{-4}$).
    In so doing, we not only demonstrate a CAPTCHA that is approximately
    four orders of magnitude more difficult to crack, but that such
    systems can be designed based on the insights gained from attack
    papers using the differences between the ways that humans and
    computers process audio.

\end{abstract}

\section{Introduction}

\captcha{}s (Completely Automated Public Turing Test to tell Computers and Humans
Apart) are a nearly ubiquitous security feature on the Web.  These
challenge-response puzzles attempt to regulate access to resources
(e.g., account sign-up, service abuse, etc) through tasks that are
simple for human beings to solve but extremely difficult for machines.
\captcha{}s now come in a wide array of formats, from the transcription
of images of character strings
\cite{baird2003pessimalprint,chew2003baffletext} and image recognition
\cite{gossweiler2009s,elson2007asirra}, to noisy audio samples
\cite{holman2007developing,sauer2008towards}. This variety benefits
inclusive computing, and ensures that users with access limitations
(e.g., vision impairments/blindness) can still use the majority of the
Web without being limited by security features~\cite{W3C}.

Unfortunately, what constitutes a challenge for computers has changed
drastically since the inception of \captcha{}s.  Nowhere has this shift
been more obvious than in audio \captcha{}s. Specifically, significant
advances in the power of neural network-driven speech-to-text systems
allow for the real-time, automated breaking of audio
\captcha{}s~\cite{bock2017uncaptcha}. However, audio \captcha{}s cannot
simply be discarded for other available methods, and in fact, remain the
fallback/supplemental methods for many other techniques. This explains why Google's re\captcha{}, one of the most popular \captcha{} services on the Internet, offers audio \captcha{} challenges (despite the availability of other methods such as math challenges, open-ended questions). Accordingly,
\textit{stronger} and more \textit{principled} techniques are required for their construction.

We argue that audio \captcha{}s are so easily breakable because of a
fundamental misunderstanding of their adversary. That is, while adding
noise to audio generally decreases comprehension by humans and
unintelligent systems, it fails to model subtle differences between the
two parties. 
The difference between how humans and computers perceive audio has been
exploited repeatedly in recent literature attacking Automatic Speech
Recognition (ASR) systems (e.g., Amazon's Alexa, Apple's Siri,
etc)~\cite{Alexa, Siri}, which allows adversaries to create starkly
different audio and transcriptions. Such approaches could potentially be
used to exploit weaknesses in speech-to-text systems to make humans and
computer observers ``hear'' dramatically different messages. 

In this paper, we leverage the large body of literature on
attacks against ASRs as a means of creating a~\textit{defense} in the
form of a robust audio \captcha{} generation algorithm. That is, instead
of simply adding noise to audio, we modify existing audio
samples in ways that significantly decrease machine transcription
accuracy without impacting human intelligibility. However, these attacks
were originally designed to force target ASRs to output specific
transcriptions for perturbed ``adversarial'' audio samples. Therefore,
their use as defenses prevents an adversary's unknown ASR from
correctly transcribing --- and therefore breaking --- the CAPTCHA. This brings
forth several fundamental questions, such as:  {\em Can these attacks be
successfully used as a defense?} If so, {\em what properties of such attacks
are required for the defense to be effective? Which among the plethora
of existing techniques are the most suitable? And what approaches (if
any) are robust to adaptive adversaries who know that the defense is in
place and can adjust their attacks accordingly?}

To answer these questions, we start our investigation by deriving
requirements for an~\textit{attack that can be used as a defense}. We
then turn to the literature and select 20 candidate attacks. Out of
three promising candidates, we find that only one, the Kenansville
attack~\cite{abdullah2019hear}, survives preliminary experiments because
of its high transferability. Transferability --- the ability of
adversarial audio samples to fool models they are not directly crafted
against --- becomes an essential property.  When an audio
\captcha{} is crafted, the defender does not know which target ASR an
adversary may eventually use for transcription.  Thus to be
maximally robust, \captcha{}s need to be mistranscribed by as many
models as possible. Our design goal is to force ASR models to output an
\textit{empty string} when presented with our audio \captcha{}s. This is
ideal because it prevents the adversary from using phonetic mapping and statistical analysis
techniques, in conjunction with incorrect or partial ASR
outputs, to help break \captcha{}s.

Further experiments reveal that the Kenansville attack is
not robust against an adaptive adversary that simply adds noise to the
CAPTCHA before transcribing it. Therefore, we propose
\kvus{}, a new algorithm that
still achieves high transferability but is robust against an adaptive
adversary. The \kvus{} defense brings two key technical innovations: (1)
the addition of Gaussian noise to the perturbed sample during CAPTCHA
creation to increase robustness, and (2) a novel thresholding technique
that clips large amplitudes of dominant frequencies in a way that
confuses ASR models but still preserves the location of the frequency
peaks so that the audio remains highly intelligible to the human ear.

We make the following contributions:
\begin{itemize}

  \item {\bf Principally-Designed \captcha{} Construction:} The design of
      audio \captcha{} has been largely ad hoc, relying on the
	experiential addition of cover noise to attempt to thwart bots.
	In contrast, our approach exploits weaknesses in modern ASR
	pipelines to intentionally target audio components that are
	ignored by our ears.

  \item {\bf Identify Best Features from Literature:} The research
      community has developed a large number of attacks on ASRs in
	recent years~\cite{schonherr2018adversarial, cai2018attacking,
	gong2017crafting, kereliuk2015deep, yuan2018commandersong,
	alzantot2018did, kreuk2018fooling}.  We categorize 20 attacks,
	point to the desirable elements found in these works, and then
	demonstrate weaknesses in attempting to directly use any of
	these schemes for \captcha{} creation in the face of an adaptive
	adversary. 
   
  \item {\bf New Attack as a Defense:}  We then design \kvus{}, a new
      defense, which prioritizes its performance for \captcha{} creation
	by maintaining the best characteristics of prior work and
	remains robust to adaptive adversaries.
    
  \item {\bf Measure Robustness, Intelligibility, and Attack Detection:}
      We implement \kvus{} and evaluate its suitability as a defense to
	produce audio \captcha{}s. We find that an attacker has a low
	chance of correctly transcribing our audio \captcha{}s ($P({\rm
	transcription}) = 4 \times 10^{-5}$). We then conduct a user
	study that shows that human transcription accuracy of \kvus{} is larger
	than even Google's popular re\captcha{}~\cite{reCAPTCH28:online} audio \captcha{} service. Finally, the
	distinctive profile of \kvus{} audio \captcha{}s provides
	ASR owners a 99.99\% probability of detecting that their ASR
	is being used to break \captcha{}s~\cite{bock2017uncaptcha}. 

\end{itemize}

We note that the nature of adversarial systems means that other
researchers may eventually break this new technique; however, the
contribution of this work is deeper than a single defense.  Rather, it is
the \textit{principle-driven} approach for designing transcription-resistant
\captcha{}s without making intelligibility worse. To assist other
researchers in testing similar schemes, we will make all of our code and
audio samples available at \url{https://AnonymizedForSubmission.com}

The remainder of this paper is organized as follows:
Section~\ref{sec:background} provides background information on audio
\captcha{}s and ASRs; Section~\ref{sec:secmod} details the threat model;
Section~\ref{sec:select_def} categorizes and evaluates the space of
attacks on ASRs and their potential use as a defense; Section~\ref{sec:kv2_details} presents the details of
the \kvus{} defense; Section~\ref{sec:kv2_eval} presents our evaluation
of the \kvus{} defense, its intelligibility, and its ability to be
detected; Section~\ref{sec:relwork} discusses related work; and
Section~\ref{sec:conc} offers concluding remarks.

\section{Background}
\label{sec:background}
\subsection{Audio \captcha{}s}
A CAPTCHA protects resources against bots by generating and
grading tests that humans can pass, but current computer
programs cannot \cite{captchasite}.  Audio CAPTCHAs are essential since
they are one of the only ways for visually impaired users to complete a CAPTCHA.
Without audio CAPTCHAs, a significant number of systems would either become
vulnerable to adversaries or lockout visually impaired users from potentially
critical services.

A good CAPTCHA must maintain accessibility to humans while protecting
the resource from adversaries. However, not
only can audio CAPTCHAs be broken by machines
\cite{bursztein2011failure,tam2008breaking,bock2017uncaptcha,solanki2017cyber}
with high success rates ($91\%$) \cite{nikolait}, but the intended human users
successfully pass them at a lower rate ($<50\%$) \cite{bigham2009evaluating}
than the attack. These CAPTCHAs are especially difficult for their intended
Visually impaired users since the audio playback can overlap with a screen reader
software, the users have to remember the audio while navigating the page
acoustically, and they are unable to compare the audio against the associated
text CAPTCHA \cite{bigham2009evaluating,bursztein2010good,dzieza_2019}. 


\subsection{Automatic Speech Recognition systems (ASRs)}
ASR systems take in speech samples and output
transcriptions of the spoken content \cite{acero1990environmental,yu2016automatic}.
They enable human-to-machine interaction, such as
issuing commands to a digital voice assistant (e.g., Amazon Alexa \cite{Alexa}).
This is not an easy task since the ASR system must be able to filter out
background noise and handle variations in accents and speech patterns.
For this reason, ASR systems' processing generally consists of three stages: preprocessing,
feature extraction, and decoding. Some models combine the feature extraction and
the decoding steps into a single Machine Learning (ML) model, and others split
the decoding step across several different ML models \cite{amodei2016deep}, but
all still perform each of these steps in some manner.

\textbf{Preprocessing:} This phase in ASR systems consists of taking in
the raw audio file and eliminating the background noise, interference, and other
information that makes it more difficult to understand the speech.  Generally, low pass filters which filters remove unwanted high-frequency noise from the
signal that are not directly related to the speech \cite{gelfand2017hearing}.

\textbf{Feature Extraction:} Next, the clean signal is converted into
overlapping frames, each of which is then passed through a feature extraction
algorithm. This algorithm retains only the salient information from each frame.
A variety of signal processing techniques \cite{rabiner1978digital} and ML
extraction layers \cite{venugopalan2014translating} have been used to establish
which features to extract. 

\textbf{Decoding:} Finally the extracted features are passed to a decoding
function, usually a machine learning model. This model takes in the extracted
features from the previous step and outputs the final transcription. There have been
a wide variety of models are used for this purpose in ASR systems, including
Convolutional Neural Networks (CNNs) \cite{abdel2012applying,sainath2013deep},
Recurrent Neural Networks (RNNs)
\cite{graves2013speech,sak2014long,sak2014sequence,sak2015fast}, Hidden Markov
Models\cite{rabiner1989tutorial}, and Gaussian Mixture Models
\cite{lamere2003design}.


\subsection{Phonemes}
Human speech is made up of various component sounds known as phonemes. Each
phoneme represents the smallest unit of sound in a word. The set of possible
phonemes is fixed due to the anatomy that is used to create them, but different
languages use different phonemes. For example, English is made up of 44
phonemes, while the International Phonetic Alphabet (IPA) contains 107
\cite{international1999handbook}. 

To perform phonetic translation, we use the Carnegie Melon University (CMU)
Pronouncing Dictionary \cite{weide2005carnegie}. This is a machine-readable
dictionary of over 130,000 words with information on their pronunciation.
Namely, it maps the words to their phonemes. The CMU Pronouncing Dictionary
outputs translations in the ARPAbet phoneme transcription code, which is
composed of 39 unique phonemes. ARPAbet is one of several phoneme transcription
codes related to the International Phonetic Alphabet (IPA), with ARPAbet
specializing in English transcription. This dictionary greatly aids feature
extraction and decoding, since it allows us to evaluate how translations are
spoken, rather than how they are written for instance, ``two bear sail'' and
``to bare sale'' are spoken identically, yet written differently.

\subsection{Breaking Audio \captcha{}s} The adversary's goal is to use automated
means to break an audio \captcha{} (i.e. transcribe it correctly). Recent
methods often involve the use of commercial ASRs for transcribing
\captcha{}s~\cite{solanki2017cyber,bock2017uncaptcha}. This method is a
multi-step process. 

The audio \captcha{} consists of sound bites or utterances of words, digits,
characters, etc. The adversary splits these into individual sounds, each
containing a single utterance. For example, the adversary splits the \captcha{}
containing the sequence ``ABMKL2'' into six individual audio files, each of
which is then passed to the ASR individually for transcription. Passing
individual utterances, instead of the entire \captcha{} audio wholesale, improves accuracy.
The adversary then parses the transcripts to ensure these are in the correct
format and only contain numbers and
digits.

In case the ASR outputs an incorrect transcript for an utterance, the adversary
uses one of two techniques to recover the real one: phonetic mapping or
statistical analysis. Phonetic mapping is employed when the ASR outputs a
phonetically similar transcript to an expected digit or character (e.g., ``too''
for the digit ``2''). 
In this case, the adversary can use phonetic mappings to identify the
digit that sounds closest to ``too'' as the correct transcript (``2'').

Statistical analysis is used when the ASR regularly outputs the same unique
incorrect transcript for a sound bite. For example, the ASR might output
``crown'' for the \captcha{} digit ``two,'' even though they are phonetically
dissimilar. An adversary with enough queries can perform simple statistical
tests to detect this pattern. As a result, she transcribes ``2'' every time the
ASR outputs ``crown.''

\subsection{Datasets} We used several audio training datasets to create our
models.  \textbf{LibriSpeech \cite{panayotov2015librispeech}:}
is a corpus of English speech used to train ASR models. The
audio is derived from audiobooks that are part of the LibriVox project. This
dataset contains about 1000 hours of text-aligned, short utterances sampled at
16kHz. This data is segmented using Smith-Waterman alignment
\cite{smith1981identification} and samples, where the audio and text do not
match, are removed. \textbf{Google Speech Commands Dataset \cite{sainath2015convolutional}:} is composed of short single-word
utterances. Specifically, the dataset consists of 65,000 one-second samples of
thousands of different people saying one of 30 words. This dataset was designed
to be used to train simple ASR models for words such as ``Yes'', ``No'', and
``Up.'' While the vocabulary is small, the number of different voices, accents,
and speech patterns in this dataset is useful in representing the many
variations of speech that exist. \textbf{LDC: ISOLET Spoken Letter Database \cite{cslu}: }is
used to train ASR models on the English Alphabet. The dataset contains two
samples of each English letter being spoken by 75 males and 75 females of
varying ages. If a letter was severely misspoken then the sample was removed
from the dataset.

\section{Problem Formulation \& Threat Model}
\label{sec:secmod}
Our goal is to design a \captcha{} generator that can perturb
benign audio samples to produce audio \captcha{}s that are intelligible to
humans, but can force the attackers' ASR into outputting incorrect
transcriptions. We also want to produce \captcha{}s that are easily detectable by
commercial ASR services, alerting when
their APIs are being misused.
 
\subsection{Adversary}
We consider an adversary whose goal is to break audio \captcha{}s (i.e., correctly transcribe their audio) in an automated way using an ASR. We assume that the adversary has the ability to query the CAPTCHA service (i.e., check if the proposed transcription is correct).

\textbf{Query Access:} \captcha{} services enforce rate
limits to prevent unfettered access from malicious users. However, we assume a
strong adversary who has no such constraints and can make unlimited queries to
the \captcha{} service.

\textbf{Success Rate:} By combining statistical analysis and phonetic
mappings, researchers have been able to break audio \captcha{}s with success
rates of up to 98.3\%~\cite{solanki2017cyber}. As in earlier works, we assume
that the adversary is successful if it can correctly transcribe the entire
\captcha{} at a success rate of just 1\%~\cite{bursztein2009decaptcha}. We show
that our \captcha{}s are robust to such a small upper bound.

\textbf{Adaptive Adversary:} This adversary has perfect knowledge of the
audio \captcha{} generation algorithm. Having this knowledge allows the
adversary to modify her behavior to overcome the algorithm, improving her
chances of breaking the \captcha{}. There are a number of ways the attacker can
modify her behavior, the two most popular of which are vulnerability
analysis~\cite{tramer2020adaptive} and adversarial
training~\cite{madry2017towards,shafahi2019adversarial,kurakin2016adversarial}.
As the name suggests, vulnerability analysis involves finding weaknesses in the
\captcha{} algorithm. The adversary can then modify the \captcha{} audio, before
passing it to the ASR, so that the ASR will output the correct transcript. On
the other hand, adversarial training involves training a local ASR to
specifically label audio \captcha{}s. The local model will be designed to be
more robust to the \captcha{} algorithm and will likely produce correct
transcripts. We show in this paper that our \captcha{}s are robust to an
adaptive adversary with perfect knowledge of our algorithm.

\subsection{Defender} Considering these adversarial capabilities, we now list
the requirements of the defender (and her \captcha{} generation algorithm) in
order to be robust to such a strong adversary.

\textbf{Intelligibility:} The defender's audio \captcha{}s
must be intelligible to humans (i.e., human
transcribable). Naively degrading audio \captcha{} quality might help prevent ASRs from
producing correct transcriptions, but will also negatively impact human
intelligibility.

\textbf{ASR Output:} Even if the ASR mistranscribes an audio sample (e.g., ``too'' instead of ``two''), an adversary can recover \captcha{} text. To prevent this, our goal is to force the ASR to output an \textit{empty string}. This is an incredibly strong threat model since ASRs can still output some text. However, by forcing an empty output, our attack makes it impractical for the adversary to recover the \captcha{} text. 

\textbf{Transferability:} An adversary can use any commercially available
ASR, or even a locally trained one, to transcribe the \captcha{}. This means
that the defender will likely not have knowledge of or even query access to the
adversary's ASR. Therefore, the \captcha{} audio must be highly transferable to force any unknown ASR to output an empty string. 

\textbf{Detection:} Adversaries leverage commercially available ASRs
to conduct attacks against \captcha{} services without the consent of the ASR
owners. These ASRs are available via API calls and handle
massive volumes of requests. Our final requirement is that the audio \captcha{} 
be identifiable by the ASR owner. This will allow the ASR owner to flag, block,
or delay the adversary's requests.

\section{Selecting Attacks as Defenses}
\label{sec:select_def} 
One way to meet the requirements outlined above is to craft
audio samples specifically designed to fool ASRs. Fortunately, there are a
plethora of attacks in the space of adversarial machine learning that try
to do exactly that~\cite{sok}. We hypothesize that these attacks can be used as a defense mechanism to
beat the ASRs used by \captcha{} breaking adversaries. In this section, we give an overview
of these attacks and evaluate them as potential defenses using our threat model.

\subsection{Different Attacks on ASRs}
\label{sec:attack_types}

\begin{table}[t]
\resizebox{\columnwidth}{!}{%
\begin{tabular}{c|c|c|c|}
\cline{2-4}
\multicolumn{1}{l|}{\textbf{}} & 
    \textbf{\begin{tabular}[c]{@{}c@{}}Potential\\ CAPTCHA\\ Use\end{tabular}} & 
    \textbf{\begin{tabular}[c]{@{}c@{}}Audio\\ Quality\end{tabular}} & 
    \textbf{\begin{tabular}[c]{@{}c@{}}Attack\\ Type\end{tabular}} \\ \hline 
\multicolumn{1}{|c|}{Taori et al. \cite{taori2018targeted}} & \xmark & Intelligible & Grad Free \\ \hline
\multicolumn{1}{|c|}{M. Azalnot et al. \cite{alzantot2018did}} & \xmark & Intelligible & Grad Free \\ \hline
\multicolumn{1}{|c|}{HVC (2) \cite{carlini2016hidden}} & \xmark & Inaudible & Misc \\ \hline
\multicolumn{1}{|c|}{Cocaine Noodles \cite{vaidya2015cocaine}} & \xmark & Inaudible & Misc \\ \hline
\multicolumn{1}{|c|}{Dolphin Attack \cite{zhang2017dolphinattack}} & \xmark & Inaudible & Misc \\ \hline
\multicolumn{1}{|c|}{Light Commands \cite{sugawara2020light}} & \xmark & Inaudible & Misc \\ \hline
\multicolumn{1}{|c|}{Roy et al. \cite{roy2018inaudible}} & \xmark & Inaudible & Misc \\ \hline
\multicolumn{1}{|c|}{HVC (1) \cite{carlini2016hidden}} & \cmark & Unintellgible & Opt \\ \hline
\multicolumn{1}{|c|}{CW \cite{carlini2018audio}} & \cmark & Intelligible & Opt \\ \hline
\multicolumn{1}{|c|}{Houdini \cite{cisse2017houdini}} & \cmark & Intelligible & Opt \\ \hline
\multicolumn{1}{|c|}{Schonherr et al. \cite{schonherr2018adversarial}} & \cmark & Intelligible & Opt \\ \hline
\multicolumn{1}{|c|}{Kreuk et al. \cite{kreuk2018fooling}} & \cmark & Intelligible & Opt \\ \hline
\multicolumn{1}{|c|}{Qin et al. \cite{qin2019imperceptible}} & \cmark & Intelligible & Opt \\ \hline
\multicolumn{1}{|c|}{Yakura et al. \cite{yakura2018robust}} & \cmark & Intelligible & Opt \\ \hline
\multicolumn{1}{|c|}{Commander Song \cite{yuan2018commandersong}} & \cmark & Intelligible & Opt \\ \hline
\multicolumn{1}{|c|}{Devil’s Whisper \cite{chen2020devil}} & \cmark & Intelligible & Opt \\ \hline
\multicolumn{1}{|c|}{Abdoli et al. \cite{abdoli2019universal}} & \cmark & Intelligible & Opt \\ \hline
\multicolumn{1}{|c|}{P-PGD \cite{PPGD}} & \cmark & Intelligible & Opt \\ \hline
\multicolumn{1}{|c|}{Kenansville Attack \cite{abdullah2019hear}} & \cmark & Intelligible & Sig Proc \\ \hline
\multicolumn{1}{|c|}{Abdullah et al. \cite{biometrics}} & \xmark & Unintellgible & Sig Proc \\ \hline
\end{tabular}}
\caption{Overview of existing attacks.
    \textbf{Inaudible:} Audio will not be heard by the human ear.
    \textbf{Unintelligible:} Audio will sound noisy to the human ear.
    \textbf{Intelligible:} Audio will be clean and understandable by the human
    ear. 
    \textbf{Grad Free:} Gradient Free, \textbf{Misc}: Miscellaneous,
    \textbf{Opt:} Optimization, \textbf{Sig-Proc:} Signal Processing
    \label{tab:overview_atks}}
\hrule
\vspace*{-5mm}
\end{table}

The goal of the \captcha{} generator is to craft audio samples that are
intelligible to humans, but can force the adversary's remote ASRs into a
a phonetically dissimilar mistranscription (i.e., force the ASR to output random garbage e.g., ``HadJSNm'' for the audio containing ``one two three'', instead of ``Juan too tree''). There are a number of attacks
that can help achieve this goal (Table~\ref{tab:overview_atks}): 

\begin{table*}[h]
\centering
\begin{tabular}{|c|c|c|c|c|}
\hline
\multirow{2}{*}{\textbf{Attack Name}} &
 \multirow{2}{*}{\textbf{Attack Type}} &
 \multicolumn{3}{c|}{\textbf{Transcription Distance}} \\ \cline{3-5} 
       &               & \textbf{Google} \cite{google_normal} & \textbf{IBM} \cite{ibm} & \textbf{Wit.ai} \cite{wit} \\ \hline
CW \cite{carlini2018audio} & Optimization   &  2.56      & 5.94     & 23.76      \\ \hline
P-PGD \cite{PPGD}    & Optimization   & 13.64      & 24.5     & 19.82      \\ \hline
Kenansville \cite{abdullah2019hear}      & Signal Processing &  49.31      & 49.76    & 48.39      \\ \hline
\end{tabular}
\caption{Transferability results. The numbers represent the Levenshtein distances between the original and the adversarial transcript. Smaller distances imply high phonetic similarity to the original, benign label (e.g., ``Juan too tree'' for the original label ``one two three''). Larger distances imply lower phonetic similarity to the original label (e.g., ``HadJSNm'' for the original label ``one two three'') Optimization attacks have higher phonetic similarity (than signal processing attacks) and therefore, can not be used as \captcha{} generators. This leaves the signal processing attack, which we further evaluate in Figure~\ref{fig:kv1_adaptive}. \label{tab:atk_eval}}
\hrule

\vspace*{-5mm}

\end{table*}

\textbf{White-Box Attacks}: 
These attacks exploit knowledge of the model's decision boundaries to craft
adversarial samples. The most popular of these are \textit{optimization} attacks.
However, it is unclear whether these attack audio can force phonetically dissimilar transferability. If so, these attacks will be a good candidate for crafting
audio \captcha{}s. There are a multitude of attacks, representative candidates
are shown in Table~\ref{tab:overview_atks}. Optimization attacks can be classified into two types: psychoacoustic and hard-clipping~\cite{PPGD}. We choose one attack from each, CW~\cite{carlini2018audio} and P-PGD~\cite{PPGD} as potential candidates. We choose these specific attacks because they have the highest rates of attack success and are architecture agnostic (i.e., can work against all ASR architectures). Other attacks in this category are too sensitive and can fail even due to minor modifications to the ASR architecture.

\textbf{Black-Box Attacks:}
These attacks do not require~\textit{any} knowledge of the target ASR to craft
adversarial samples. The two most popular types are \textit{gradient-free} and
\textit{signal processing} attacks. Gradient-free attacks are a constrained
version of the optimization attack family because they estimate the decision
boundaries of the target ASR. However, these
produce lower quality adversarial audio and therefore, are not good
candidates for audio \captcha{}.

On the other hand, signal processing attacks exploit the imperfections in the signal processing
algorithms used in the ASR pipeline that are designed to emulate the human
ear~\cite{biometrics}. These attacks can produce
samples that can successfully transfer to and exploit remote models. This makes
signal processing attacks good candidates for generating audio \captcha{}s. Of
the signal processing attacks, only \kvog{} can produce intelligible
audio~\cite{abdullah2019hear}. As a result, we use it as a candidate attack
for our experiments.

\subsection{Evaluation: Transferability}
Having chosen the three attack candidates~\cite{PPGD,carlini2018audio,abdullah2019hear}, we evaluate these as potential \captcha{} algorithms. The goal is to
quantify how well these attacks provide \textit{phonetically dissimilar} transferability. Attack audio
with higher transferability will have a higher probability of fooling the
adversary's ASR.

\textbf{ASRs:}
For this experiment, we chose a total of four models~\cite{uberi}: one surrogate
(DeepSpeech~\cite{ds1_tf}), and three remote (Wit.ai~\cite{wit},
Google~\cite{google_normal}, and IBM~\cite{ibm}). The surrogate model is a local model available to the \captcha{} service owner. This is designed to emulate the adversary's ASR or the remote ASR. We chose the popular and open-source DeepSpeech ASR as the surrogate. In each case, the candidate algorithm constructs adversarial samples for the surrogate ASR and transfers them to the remaining remote ones.

\textbf{Dataset:}
We followed the procedure laid out in previous works~\cite{sok,PPGD}. We pooled
a set of 1000 benign audio files by randomly sampling from the
LibriSpeech~\cite{panayotov2015librispeech} dataset. Each of these
was transcribed using the surrogate DeepSpeech model to ensure an average Word
Error Rate of less than 0.1.

\textbf{Adversarial Audio Generation:}
Each of the 1000 audio files was then perturbed using the three candidate
attacks to work against the surrogate DeepSpeech model. Each audio sample was perturbed to random target transcripts using the three attacks using 1000 attack
iterations. In the end, we have 1000 audio samples from each
attack.

\textbf{Transferability:}
Next, we pass each of the 1000 attack audio samples to the three remote ASRs and
retrieve the corresponding adversarial transcriptions. We calculate the phonetic similarity using the Levenshtein distance scores between the phonetic representations (extracted using the CMU Pronouncing Dictionary~\cite{cmu}) of the original transcript and the one produced by the remote ASR.


\textbf{Results:}
Table~\ref{tab:atk_eval} shows the results of our experiment. Smaller distance
scores mean that the adversarial transcript (produced by the
remote ASR) is phonetically similar to the original label. We can observe that optimization attacks have lower distance scores
compared to the signal processing attack. This demonstrates that optimization
attack \captcha{}s will produce transcripts similar to the original CAPTCHA label (e.g., ``too'' for the number ``two''), allowing the adversary to easily reconstruct CAPTCHA text. Thus, optimization attacks are unfit for use as \captcha{}
generation algorithms. In contrast, the \kvog{} attack resulted in the
highest distance scores. This indicates that the \kvog{} attack fulfills the
transferability requirement for \captcha{} generation mechanism.

\subsection{Evaluation: Adaptive Adversary}
As described in our threat model, any valid \captcha{} generation method should succeed in the presence of an adaptive
adversary. Having
filtered the potential attack candidates to just \kvog, we now evaluate its
performance against an adaptive adversary. This adversary can either perform adversarial training or vulnerability analysis.
While the original paper demonstrated that \kvog{} is robust to adversarial
training (reasons for which discuss in the next section), in this paper we
perform vulnerability analysis.

\textbf{Attack Steps:}
We first discuss how the
\kvog{} attack works. The pipeline performs signal
decomposition using the DFT on a given audio sample. Next, all the frequency bins with power less than
the threshold are set to zero, reconstructed back into a raw signal, and
passed to the ASR. If the ASR
outputs the wrong transcript, we lower the threshold and start again. This is
because smaller thresholds result in better audio quality. If the ASR outputs the correct transcript,
the threshold is increased and the steps are repeated. A diagram of this
attack is available in Appendix~\ref{sec:app} (Figure~\ref{fig:baseline}).


\textbf{Why The Attack Works}: 
This attack exploits the differences between how humans and ASR pipelines
process audio. Specifically, the attack discards low
power frequencies (by setting their intensities to zero). Due to their already
low intensity, these frequencies are inaudible to the human ear. Therefore,
removing them does not affect the audio quality for human listeners. However, passing
these perturbed audio samples to the ASR forces an incorrect output. That is
because audio samples with zero intensity frequencies do not occur in the
natural world. As a consequence, such samples do not exist in the training
datasets for ASRs and therefore force mistranscriptions.



\textbf{Initial Hypotheses:}
Audio samples with empty frequency bins do not exist in the natural world, and therefore not in the ASR's training dataset. As a consequence, these samples force ASRs into producing a mistranscription. Adding
power to the empty frequency bins will undo the effects of the attack and allow the ASR to produce the correct output.

\textbf{Methodology:}
The adaptive adversary will use this knowledge to add small intensities of
power \textit{back} to the spectrum. One way to do this is to add Gaussian
(white) noise. Using any other noise might bias some frequency bins
over others. In contrast, Gaussian noise increases the power evenly across the previously empty frequency bins,
thereby undoing the effects of the attack. To account for the noise, the attack
algorithm will need to perturb the audio even more by setting a higher threshold.
This will further degrade the quality of the attack audio but will make it robust to
the Gaussian noise. In essence, the attack will balance
robustness to Gaussian noise against the intelligibility of the attack audio
sample. 

In this experiment, we quantify the intelligibility of the \kvog{}
attack in the presence of this adaptive adversary. Specifically, we use multiple magnitudes of noise to produce different noised versions of the adversarial audio. The attack succeeds \textit{only} if all the resulting noised versions produce an empty string. We also run a similar baseline experiment, without the presence of the adaptive adversary. For a fair comparison, we maximize \kvog{}’s intelligibility and use binary search to find the optimal perturbation parameter (as recommended by the original authors).

\textbf{Setup:}
For this experiment, we will opt for datasets that are most frequently used in real-world \captcha{} services\footnote{ 
We have placed these findings on current \captcha{} services in the Appendix Table~\ref{tab:captcha_service_overview}.}, consisting of words, letters, and digits. Our dataset included 26 English
language characters from the LDC dataset~\cite{cslu}, and 10 digits, and 20
words from the Google speech commands dataset~\cite{warden2018speech}. We then
randomly sample 10 audio files from each of these labels.\footnote{We only chose
10 samples per label because we make 1,380 queries for each sample ballooning
to a total of 900,000 queries. Increasing the total labels to greater than 10
would have made it difficult to use any of the commercial ASRs since these
impose strict rate limiting.} We only use files that the commercial ASRs
transcribe correctly. This results in approximately 560 audio files for the
adaptive adversary and the baseline experiments.

We use Gaussian noise with the zero-mean, and variance selected on a
logarithmic sweep between 0.001\% to 20\% of the maximum amplitude found in the
attack audio, resulting in 46 amplitudes. We generate five different Gaussian
noise samples for each of these 46, producing 230 noise samples. We repeat the trial five times since Gaussian noise is random, and could
potentially impact our results. Each of the resulting 230 noise signals is
added to the perturbed audio sample, resulting in 230 noise outputs. These are
then passed to the model for transcription. 

We pick four commercially available ASRs that the adversary can employ. These
included Microsoft Azure, Google Speech, IBM, and Wit.ai.

\begin{figure}
 \includegraphics[width=\linewidth]{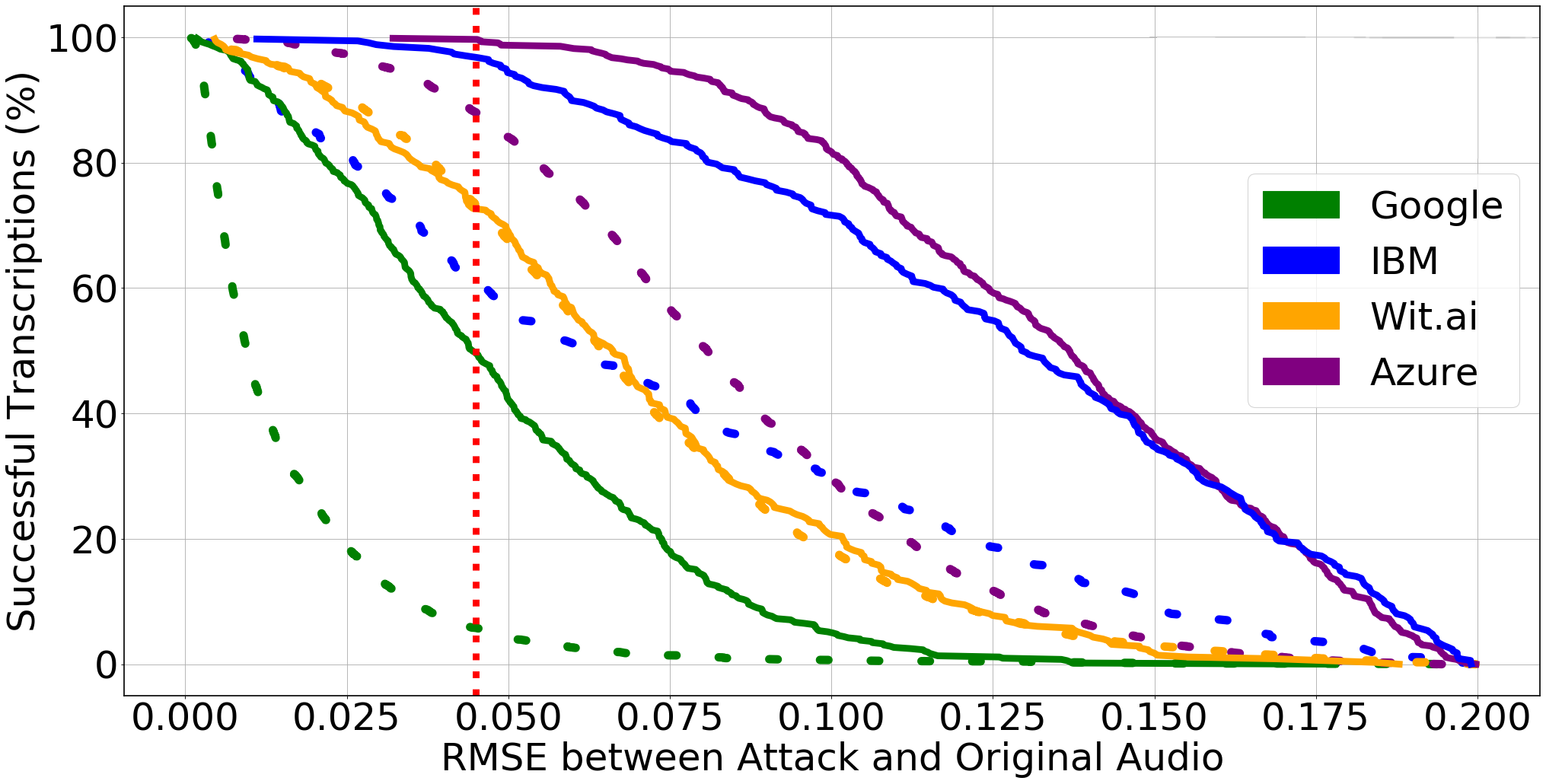}
 \caption{We evaluate \kvog{} against four popular ASRs. Each line shows the change in ASR accuracy with respect to \kvog{}'s distortion (measured using the RMSE). The vertical red line is the maximum distortion before which the audio starts to become unintelligible. The dotted lines represent the normal ASR, while the solid ones represent the ASR controlled by an adaptive adversary. \kvog{} requires significantly more distortion to fool the adaptive adversary's ASR. Therefore, the solid lines have a dramatically higher RMSE than the corresponding dotted ones. The increased distortion makes the audio unintelligible. Therefore, \kvog{} audio can not be used in \captcha{}s.
}

 \label{fig:kv1_adaptive}
\vspace*{-5mm}
\end{figure}

\textbf{Results:}
We evaluate the effectiveness of the adaptive adversary similar to
the authors of the original \kvog{} paper. Specifically, we compare the
transcription success rate of several ASRs against the acoustic distortion
measured using the Root Means Squared Error (RMSE) between the original and
perturbed audio. Higher RMSE means greater audio distortion to successfully fool the model. 

Figure~\ref{fig:kv1_adaptive} shows the results of these experiments. As the
RMSE increases, the successful transcriptions decrease. And the faster the rate of decrease (greater steepness of line), the better the attack.
We can observe that the steepness decreases markedly for the adaptive adversary (solid line), than without one (dotted line). For example, consider the results for the
Google ASR (green). The percentage of successful transcripts degrades far more
slowly for the adaptive
adversary (solid green line) than the baseline (dotted green line). This indicates that the presence of the adaptive adversary (green solid line) results in higher audio distortion, significantly degrading human audio intelligibility.

To better visualize the intelligibility of the audio, we denote the GSM audio codec's
average RMSE as the baseline for audio comprehension (the red vertical dashed
line in Figure~\ref{fig:kv1_adaptive}). The GSM audio codec is used during 2G
cellular calls, which is the most common global means of telephony
communication. We assume any audio created that has an RMSE greater than
GSM (to the right of the GSM line in Figure~\ref{fig:kv1_adaptive}) is unintelligible, and thus
unfit as an audio \captcha{}. 

Examining the Google results again, we can see that the presence of the adaptive
adversary results in a significant degradation in the attack audio quality. In
the baseline, \kvog{} can force more than 90\% of the audio samples to
mistranscribe, before crossing the GSM line, and become unintelligible.
However, in the presence of the adaptive adversary, \kvog{} is only able to
degrade the transcription success to 50\% before the audio becomes too degraded
for \captcha{} use. We see these results consistently across all models (except
for Wit.ai). These plots, therefore, show that the \kvog{} attack is
insufficient for \captcha{} generation as it is unable to produce intelligible
audio in the presence of an adaptive adversary.

\textbf{Take Away:} Despite the plethora of candidate attacks, none of them can
be used as \captcha{} generators. Attacks either fail to transfer between
different ASRs or are vulnerable to adaptive adversaries. Thus, existing work
does not suffice. Lastly, even though we use RMSE in this experiment, such metrics are a poor
measure for human audio intelligibility~\cite{biometrics}. Therefore, in
later sections, we evaluate the \captcha{} quality under the adaptive
adversary via a user study.

\section{\kvus{} Overview}\label{sec:kv2_details}
So far, we have demonstrated that existing works from adversarial ML can not be used as \captcha{} generation algorithms.  The three potential candidate \captcha{} generators do not meet the threat model requirements: The two optimization attacks failed to exploit remote ASRs and the
\kvog{} attack is vulnerable to adaptive adversaries. Based on the
lessons learned from our experiments, we design a new strategy to overcome the
limitations inherent to prior work. 

We design our \captcha{} algorithm, \kvus{}, based on the signal processing family of attacks. This is primarily because this family of algorithms has high transferability against remote models (a necessary condition for CAPTCHA audio), as we saw in the last section. However, crucially, these attacks (including Kenansville) lack robustness to adaptive adversaries. Our new \kvus{} defense overcomes this limitation by exploiting the differences in how humans and ASRs process audio, and produces \captcha{} that are robust to bots.

\subsection{Defense Steps}

Our defense has three major steps: decimation, clipping, and noising. The Kenansville method has none of these steps, which explains its lack of robustness to adaptive attackers. Each of our steps is either motivated by the requirements of the threat model or the
lessons learned during the evaluations in the previous section. 


\begin{figure}
 \includegraphics[width=\linewidth]{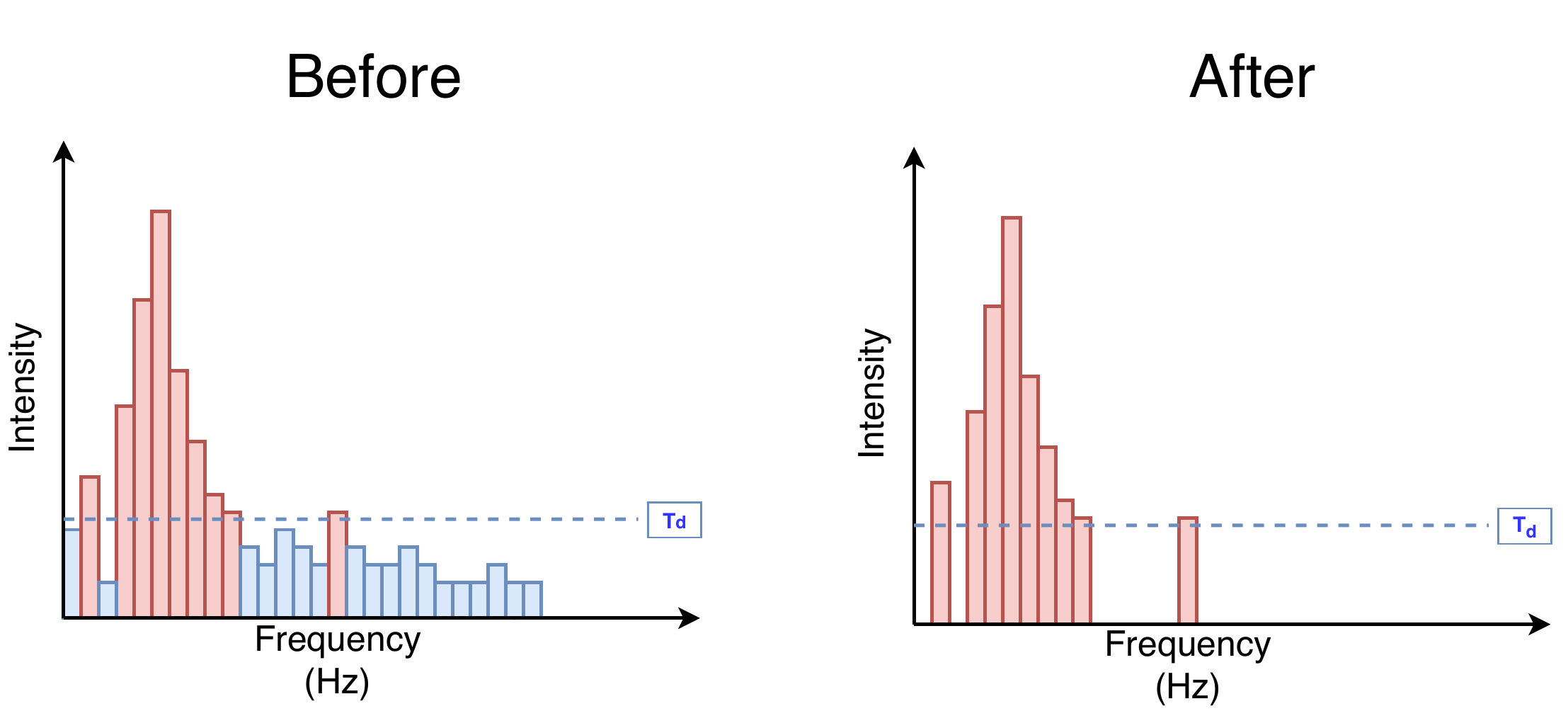}
 \caption{From the \kvus{} pipeline: Decimation is performed on the audio
    spectrum. We set all the frequency bins (blue) below
    $T_d$ to zero.}
 \label{fig:decimation}
\vspace*{-5mm}
\end{figure}
\textbf{Step 1: Decimation:}
Decimation is the process of discarding frequencies below a predefined threshold $T_d$. This property is known to be robust against adversarial training~\cite{abdullah2019hear}, as it prevent the ASR from converging during training. As a result, models trained on decimated audio have lower accuracy than their baseline counterparts. Intuitively, it is because  frequency bins are dissimilarly discarded across the spectrum (even for two utterances of the same word), which affects the model's ability to learn proper decision boundaries. We use this key observation to design the first step of our defense, shown in Figure~\ref{fig:decimation}. Here, we take the DFT of the audio sample, discard the frequencies below the threshold $T_d$ (shown in blue). We then use the remaining frequencies (shown in red) to reconstruct the raw audio waveform.

\begin{figure}
 \includegraphics[width=\linewidth]{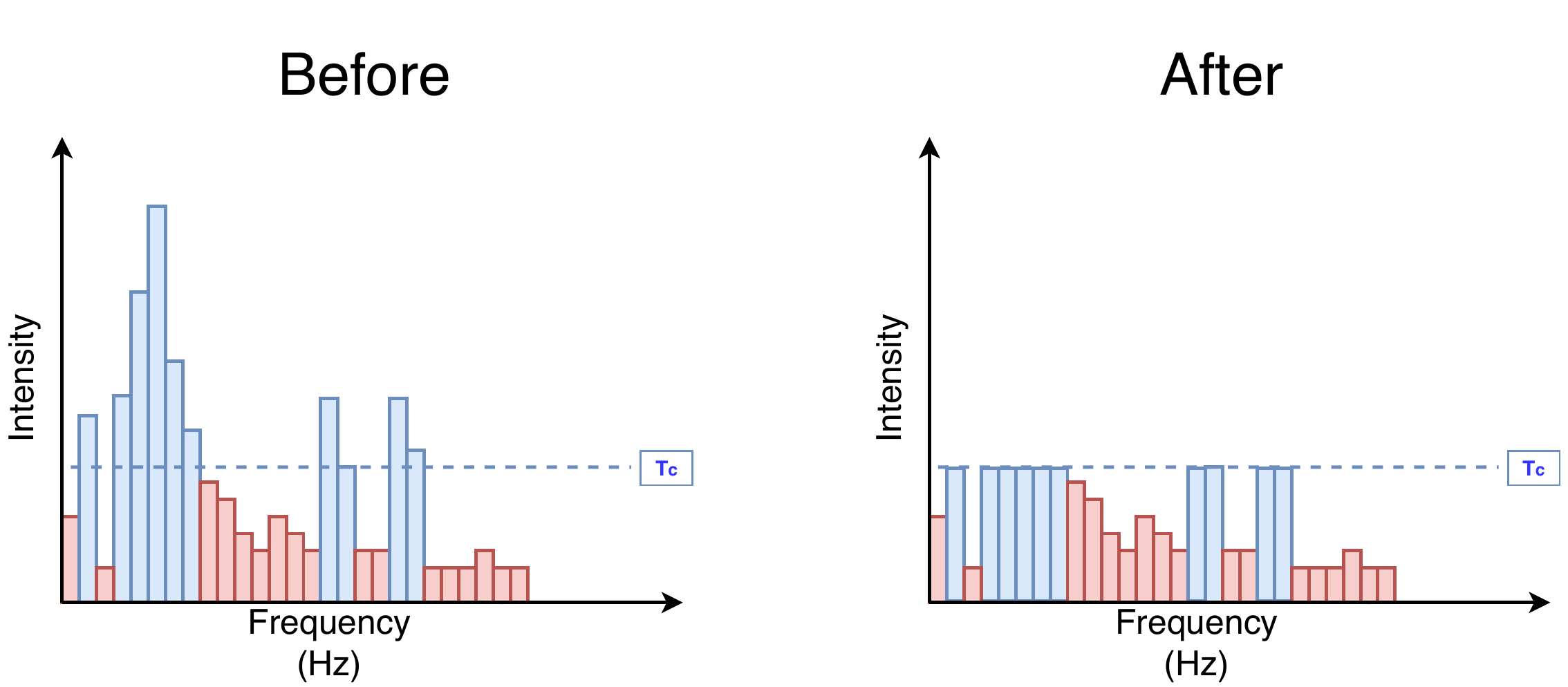}
 \caption{From the \kvus{} pipeline: Clipping is performed on the audio
    spectrum. We set all the frequency bins (blue) above the $T_c$ to the value
    of $T_c$.}
 \label{fig:clip}
\vspace*{-5mm}
\end{figure} 
\textbf{Step 2: Clipping:} We use spectral clipping to force the model to output an empty string. The human ear largely relies on small bands of dominant frequencies (e.g., formants in the case of
vowels) to identify individual phonemes~\cite{stevens2000acoustic}. For example, the phoneme /\ae/ (such as in the word ``h\underline{a}t'') has an
average first formant of 585 Hz and the second formant of 1710 Hz. These frequency peaks vary between people and within the same speaker, thus making the raw frequency values less critical to intelligibility. Instead, humans rely on the ratio between these two frequency peaks. In the case of /\ae/, the difference between the first two peaks is generally around 1.9 times the value of the first formant~\cite{brend1990practical}. The human ear can still identify phonemes even if the maximum amplitude within
these peaks have been clipped, as long as the location
of the frequency bands with the spectrum remain unchanged and the frequency
band still dominates its neighboring frequencies.

In the following procedure, we propose clipping the dominant frequencies of the
spectrogram as a means of fooling the model. Clipping these frequencies will
create phonetic structures that do not exist in the natural world, and will
therefore fool the ASRs. However, since clipping does not change the location of
the dominant frequencies, it will still sound the same to the human ear.

Additionally, clipping has the added benefit of being robust to 
the Gaussian noise-based adaptive adversary. By clipping,
we are leveling out the structure of the peaks that the ASR needs for correct
transcriptions. Adding random noise will not recreate the structure of the
peaks that we clipped out and therefore, not result in the correct original transcript.

For this step, we will use a clip function to clip the values outside a specified interval [0, $T_c$] (Figure~\ref{fig:clip}), calculated using the function:
\begin{equation} 
T_c = \max{(DFT(x))}-\alpha
\end{equation}
where $\alpha$ is the clipping value between [0, $\max{(DFT(x))}$]. All the frequencies that have higher intensity
than a threshold $T_c$ (shown in blue) will be clipped to $T_c$. All the frequencies that lie below the $T_c$ (shown in red) will remain unchanged. Therefore, higher values of $T_c$ will clip out more components, producing lower-quality audio. Such clipping will impact the dominant frequencies (i.e., the ones with the highest power) while, maintaining the overall phonetic structure will. We use a binary search to efficiently locate the minimum $T_c$ to minimize the impact on audio
quality.

\textbf{Step 3: Noising:}
As seen in the previous section, an adaptive adversary can break \captcha{}s by adding small amounts of noise, enabling the ASR to transcribe the audio
\captcha{} correctly. Therefore, we account for the noising step as part of our defense pipeline,  We will add noise to every sample produced with our defense before we pass it to the ASR. This will ensure that we take account of the adaptive adversary during audio \captcha{} generation.


\begin{figure}[!t]
 \includegraphics[width=\linewidth]{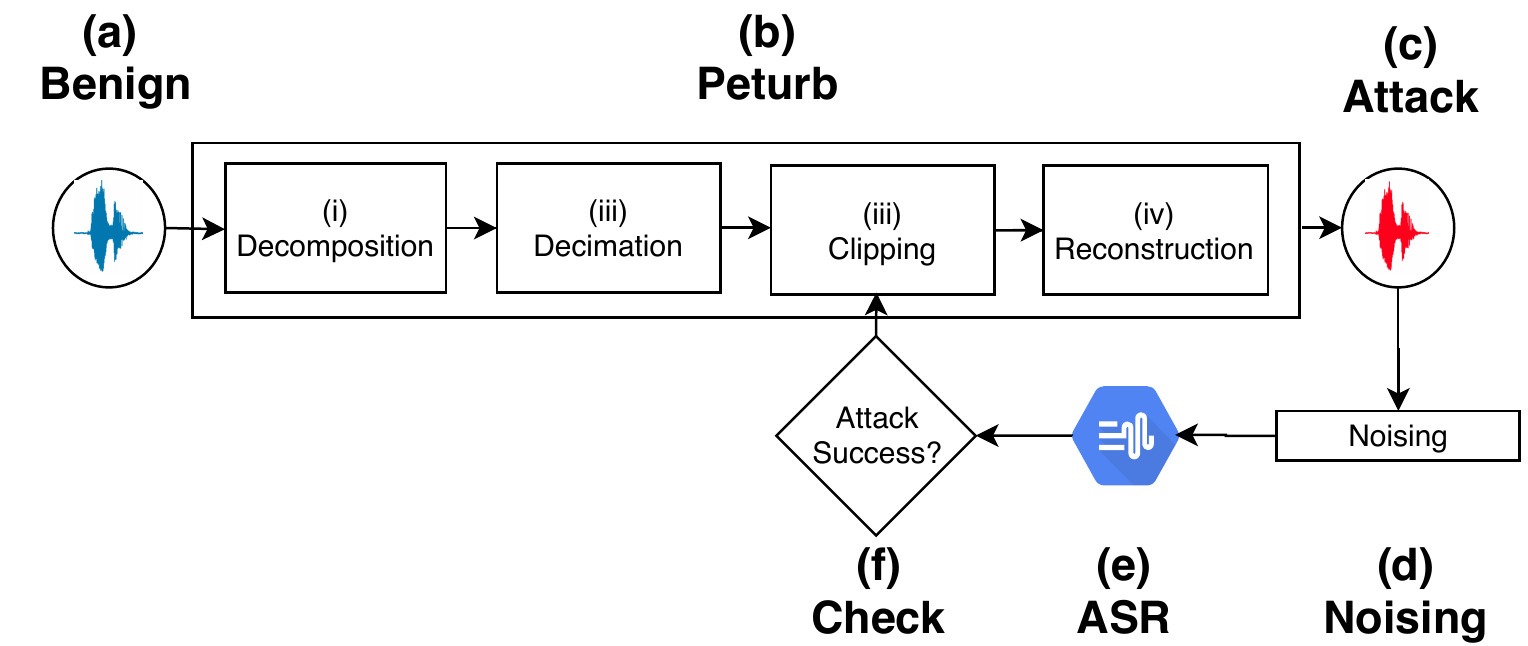}
    \caption{Full \kvus{} pipeline. (a) The audio sample is (bi) is
    decomposed into its frequency components. (b.ii) Decimated by discarding low-intensity frequencies. (b.iii) Clip the
    decimated frequencies based on a threshold. (b.iv) Reconstruct the modified
    spectrum into a (c) raw audio sample. (d) Add noise to the
    audio, to account for the adaptive adversary (e) and pass it to the ASR. (f) Check for success. If so, lower the
    clipping threshold and rerun the steps.}
 \label{fig:kv2}
\vspace*{-3mm}
\end{figure}

The full pipeline is shown in Figure~\ref{fig:kv2}. We perturb an audio
sample (a), by first performing signal decomposition (bi). We then decimate (bii) the audio by some fixed threshold,
$T_d$. Next, we clip the spectrum (biii) using a variable threshold, $T_c$,
and then reconstruct the raw audio waveform. To account for the adaptive
adversary, we noise the audio (d) by adding Gaussian noise, before passing it to
the ASR for transcription. We then check whether the ASR output an empty string.
If so, we reduce $T_c$ and repeat the steps until we find the lowest value for
$T_c$ that forces the ASR to produce an empty string.

\subsection{Differences between \kvog{} and \kvus{}:}
Though both \kvus{} and \kvog{} belong to the same broad family of attacks (i.e., signal processing attacks), they are starkly different. \kvus{} involves a three step process (decimation, clipping, and noising) that is missing in \kvog{} (which only performs decimation). This enables \kvus{} to force an ASR to output an empty string, while still maintaining high audio quality. In contrast, \kvog{} only produces mistranscriptions, some of which are phonetically similar to the original \captcha{} label.  As we show in the next section, \kvus{}’s three step design results in much higher human intelligibility, while ensuring greater robustness.
\begin{table}[!t]
\centering
\begin{tabular}{c|c|c|c|c|}
\cline{2-5}
 & \begin{tabular}[c]{@{}c@{}} \textbf{Yeehaw} \end{tabular} & \begin{tabular}[c]{@{}c@{}} \textbf{Kenansville} \\ \cite{abdullah2019hear}\end{tabular}
 & 
 \begin{tabular}[c]{@{}c@{}} \textbf{CW} \\ \cite{carlini2018audio}\end{tabular} &
 \begin{tabular}[c]{@{}c@{}} \textbf{PPGD}\\ \cite{PPGD}\end{tabular}  \\ \hline
\multicolumn{1}{|c|}{\textbf{Intelligibility}} & \cmark & \cmark & \cmark & \xmark \\ \hline
\multicolumn{1}{|c|}{\textbf{Transferability}} & \cmark & \cmark & \xmark & \xmark \\ \hline
\multicolumn{1}{|c|}{\textbf{\begin{tabular}[c]{@{}c@{}}Adp-Adv\\ (Adv-Training)\end{tabular}}} & \cmark & \cmark & N/A & N/A \\ \hline
\multicolumn{1}{|c|}{\textbf{\begin{tabular}[c]{@{}c@{}}Adp-Adv\\ (Vuln-Analysis)\end{tabular}}} & \cmark & \xmark & N/A & N/A \\ \hline
\multicolumn{1}{|c|}{\textbf{Detection}} & \cmark & N/A & N/A & N/A \\ \hline
\end{tabular}

\caption{Comparison between our \captcha{} algorithm and existing ones using the
    threat model. Our
    algorithm meets all requirements. \cmark: Meets Requirement. \xmark: Does not
    meet requirements. N/A: We did not test since other requirements not met.
    \label{tab:kv2_eval}}
\hrule
\vspace*{-5mm}
\end{table}
\section{Evaluation of \kvus{}}\label{sec:kv2_eval}

\label{sec:kv2_eval}
We now evaluate \kvus{} against the
threat model (Table~\ref{tab:kv2_eval}) and compare it against other \captcha{} algorithms from literature and industry.

\subsection{Transferability}

As discussed in Section~\ref{sec:secmod}, audio \captcha{}s should be able to
transferable to unknown remote ASRs (i.e., force the ASR to output an empty string). This prevents the attacker from using phonetic
mapping or statistical analyses to get the extract transcript. In this
experiment, we will quantify our attack's transferability. We craft audio \captcha{}s for each (surrogate) ASR using our \captcha{} algorithm and
then pass these to the remaining remote models to calculate the transferability rates.

\textbf{Setup:} We use \kvus{} to perturb each audio file. The \captcha{}
quality controlled using two parameters: the decimation ($T_d$) and
clipping ($T_c$) thresholds. We use binary search to find the values of these parameters.
{\it We will continually perturb the audio for different values until the surrogate model produces an
empty transcript.} We target four models: Google, Wit.ai, IBM,
and Azure. 

\textbf{Dataset:} Similar to the previous experiments, our audio dataset
consists of words, numbers, and characters. We will pool benign audio samples
from across 56 labels (20 words, 10 digits, and 26 characters). We will
10 audio samples for each file, resulting in 560 utterances.


\textbf{Results:}
Table~\ref{Tab:kv2_transferability} shows the results of the transferability
experiments. Samples created for the Azure model are the most transferable. This
means that \captcha{}s designed to evade Azure (i.e., force empty string output)
will also evade remote models as well. In the worst case, the
audio generated against Azure will have an evasion rate of at least 81\% against
remote models. Upon further inspection, we observed that Azure required the highest $T_c$ of all the models. Therefore, crafting 
audio samples for Azure will guarantee success against a six-length \captcha{} with at most probability of $P({\rm transcription}) = (1-0.81)^6 = 4\times10^{-5}$.
This is orders of magnitude below the 1\% success rate needed for the
adversary to be considered successful.

\begin{table}[t!]
\centering
\begin{tabular}{cc|c|c|c|c|}
\cline{3-6}
 & & \multicolumn{4}{c|}{\textbf{Remote}} \\ \cline{3-6} 
 & & \multicolumn{1}{c|}{\textbf{Azure}} & \multicolumn{1}{c|}{\textbf{Google} } & \multicolumn{1}{c|}{\textbf{IBM} } & \multicolumn{1}{c|}{\textbf{Wit.ai} } \\ \hline
\multicolumn{1}{|c|}{\multirow{4}{*}{\rotatebox[origin=c]{90}{\textbf{Surrogate}}}} & \textbf{Azure} & 100\% & ~\textbf{93\%} & ~\textbf{97\%} & ~\textbf{81\%} \\ \cline{2-6} 
\multicolumn{1}{|c|}{} & \textbf{Google} & 54\% & 100\% & ~\textbf{96\%} & 76\% \\ \cline{2-6} 
\multicolumn{1}{|c|}{} & \textbf{IBM} & 31\% & ~\textbf{82\%} & 100\% & 64\% \\ \cline{2-6} 
\multicolumn{1}{|c|}{} & \textbf{Wit.ai} & 43\% & 76\% & ~\textbf{97}\% & 100\% \\ \hline
\end{tabular}
\caption{Transferability experiments for \kvus{} show that it has high transferability. Numbers in~\textbf{bold} show the highest levels. Models names have been arranged in descending order of their
    transferability rates.\label{Tab:kv2_transferability}}
\hrule
\vspace{-5mm}
\end{table}

\subsection{Intelligibility}
\label{sec:user_study}
So far, we have demonstrated the robustness of \kvus{} to remote ASRs. We want
to ensure that our audio is intelligible to humans. As a result, we conduct a user study to examine
intelligibility. We compare the human transcription rate of \kvus{} against and re\captcha{} (which
is \textit{the} most popular \captcha{} service~\cite{recaptcha_pop}). Our goal is to demonstrate that \kvus{} is as
intelligible as other services while providing robustness to ASRs. It is important to note that the goal of this paper is not to
develop the most~\textit{usable} \captcha{}s ( like previous
works~\cite{fanelle2020}). Instead, we want to develop the
most~\textit{intelligible} \captcha{}s. We can improve overall usability with
findings from other work~\cite{fanelle2020}.
 
We also include in our experiment \kvog{} audio samples that can fool an adaptive adversary. While we have shown in earlier experiments that \kvog{} produces audio samples with high distortion when trying to evade an ASR controlled by an adaptive adversary, this experiment will demonstrate that these audio samples are not intelligible to humans listeners. This will justify the need for \kvus{}.


\textbf{Setup:} 
We conducted a single-session, within-subject user study (IRB reviewed and exempted), with participants
recruited from Amazon's Mechanical Turk (MTurk) crowd-sourcing platform.
Participants were located within the United States, had an approval rating for
Human Intelligence Task completion of more than $95\%$, and had ages from $18$
to $55$ years. During the study, each transcribed three randomly selected \captcha{}s.

Four \captcha{}s were sampled from each of \kvus{}, \kvog{} (Adaptive Adversary), and reCAPTCHA. Each audio samples from re\captcha{}, \kvus{}, and \kvog{} contain either three or four-word phrases from the English language. Each participant was presented with three CAPTCHA audio samples from the
three CAPTCHA generators in random order, and the participant was asked to transcribe. Transcription accuracy was evaluated using the word level edit distance. We treat each word as an individual token (instead of each character) as a misspelled word is considered a failure by the \captcha{} service.

In addition to transcribing, the participants were asked the number of times they played back the audio, a demographic questionnaire, and compensated with \$1 for taking part in the
study. Before running the experiment, we calculated the sample participant size
to be 199, with an effect size of $0.5$, a type-I error rate of $0.05$, and
statistical power of $0.8$ and recruited 201 participants. On average, participants took three minutes to complete the study.

 Lastly, we passed the \captcha{} samples to the Google Speech API to see which ones
transcribe correctly. If the model did not transcribe
them correctly, we use the un\captcha{}~\cite{bock2017uncaptcha} strategy of
passing one utterance at a time. If the model got a correct transcription in any one of the two scenarios, that \captcha{} audio was considered broken.

\textbf{Results:}
\label{user-study-result} 
Table~\ref{tab:user_study_results_main} shows the results of our experiment. There are three main takeaways here. First, \kvus{} audio is more intelligible than re\captcha{}. Smaller edit distance scores means smaller transcription errors. Table~\ref{tab:user_study_results_main} shows that \kvus{} has the smallest score. After running a repeated-measure t-test, we found that transcription errors (missing or misspelled words) for \kvus~ are significantly lower than both reCAPTCHA~($t = 2.17, p < 0.05$) and \kvog~($t = 11.63, p < 0.001$). Table~\ref{tab:user_study_results_main} presents summary statistics (mean and standard deviation) of transcription evaluation based on word-level edit distance. 

Second, the \kvog{} produces highly unintelligible audio. We can see that \kvog{} has the highest distance score of 3.09 across all the \captcha{} samples. Upon further inspection, we noted that over 55\% of the users had left the \kvog{} audio transcriptions blank as they could not understand
the audio at all. In stark contrast, no transcription was left blank for \kvus{}. This proves that \kvog{} audio is unintelligible and can not be used for audio \captcha{}s. Besides, participants, on average, had to playback \kvog{} three times for transcription, while it took them two playback attempts to transcribe \kvus{} and re\captcha{}. This also supports the lower audio quality
and incomprehensible nature of \kvog{}. Even though \kvog{} and \kvus{} belong to the same family of signal processing attacks, they have widely different characteristics and audio quality.

Third, Table~\ref{tab:user_study_results_main} shows that re\captcha{} is broken as its samples were correctly transcribed by the ASR. This was possible even without using the stronger un\captcha{} strategy. In contrast, the ASR did not correctly transcribe any of the audio samples from
\kvus{} or \kvog{}, even when using the un\captcha{}
method. This means that even a lazy adversary can break re\captcha{} by passing audio to the ASR wholesale. In contrast, such an adversary
is unable to break \kvus{} and \kvog{}. These results confirm our hypothesis that \kvus{} is a viable \captcha{}
generator. Our defense has both~\textit{higher} transcription accuracy and robustness than the popular re\captcha{}.

\begin{table}[t!]
	\centering
	
	\begin{tabular}{|c|c|c|}
		\hline
		\textbf{\begin{tabular}[c]{@{}c@{}}\captcha{}\\ Algorithm\end{tabular}} & \textbf{Broken}  &\textbf{\begin{tabular}[c]{@{}c@{}} Edit Distance  \\ (M/SD)\end{tabular}}\\ \hline
		
		reCAPTCHA~\cite{von2008recaptcha}& \cmark & $1.50/1.49$  \\ \hline
		\textbf{\kvus} & & $1.21/1.69$  \\ \hline
		\begin{tabular}[c]{@{}c@{}}Kenansville~\cite{abdullah2019hear}\\ (Adaptive Adversary)\end{tabular} 
		& & $3.09/1.45$  \\ \hline
	\end{tabular}
	
	\caption{User study results. First, our \kvus{} defense is robust to ASRs, while having the lowest edit distance scores (even lower than re\captcha). Second, \kvog{} has the highest scores, meaning its audio samples are least intelligible. Third, re\captcha{} is broken (i.e., is unable to fool ASRs).  \textbf{M:} Mean, \textbf{SD:} Standard Deviation. \label{tab:user_study_results_main}}
	\hrule
	\vspace*{-5mm}
\end{table}

\subsection{Adversarial Training}\label{sec:adv_training}

So far, we have demonstrated that our defense can produce intelligible audio that
can fool commercial ASRs with very high success rates. This effectiveness can be attributed to two reasons. First, commercial ASRs are not designed
to break \captcha{}s. Second, their training data does not account for our
\kvus{} audio. To overcome these limitations, an adaptive adversary can train a local model specifically designed to
break our \captcha{}s. This is known as adversarial training and has been shown to be the most robust means of tackling adversarial samples~\cite{tramer2020adaptive}. However, in this experiment, we show
that even under ideal circumstances, adversarial training will be unable to
break our \captcha{}s.

As with any model, the model trained by the
adversary will be more accurate on certain labels than others. We can assume,
however, that the defender will also have this knowledge since she too
can train a local model. The defender can use this information to decide how
frequently certain labels should be used in \captcha{}s. For example, if the
defender knows that the adversarial model has poor accuracy on ``2,'' she can
have this label occur more frequently within the \captcha{}s she produces.
Therefore, to calculate the accuracy of the adversary's local model, we can not
simply take the accuracy of the test set. We instead need to weigh the samples
by how frequently a defender will select them for use in a \captcha{}. 


\textbf{Methodology:} We exactly follow the adversarial training methodology described in prior works~\cite{abdullah2019hear}. We perturb the~\textit{entire} dataset (of words, letters, and digits) using \kvus. We use all the samples from the datasets to ensure high model accuracy, which is a total of 72,800 audio samples. Instead of merely augmenting the data, perturb all the samples from both the train and test sets. This ASR will be specifically built for breaking \captcha{} samples produced using \kvus{} (instead of being used for general purpose speech recognition). This will ensure greater effectiveness at breaking \captcha{}s as the decision boundaries will not be skewed towards some other recognition task. Additionally, since the attack is non-stochastic (i.e., does not~\textit{randomly} decimate or clips frequency bins), perturbing one sample for each attack parameter is enough~\cite{abdullah2019hear}.
We then conduct two sets of experiments to observe the effect of changing $T_d$ and $T_c$ on model accuracy.

\textbf{Models:} Prior work has shown that certain ASR architectures are particularly robust to adversarial audio~\cite{dlt}. Factors such as (1) presence of the RNN layer (e.g., LSTMS or GRUs), (2) multi-sequence output, and (3) large model complexity~\cite{dlt} can decrease robustness to adversarial samples, and therefore, prevent an ASR from breaking \captcha{}s. This rules out the possibility of using any general-purpose ASR, such as DeepSpeech, for adversarial training (e.g., via fine-tuning, transfer learning, etc.). This is because DeepSpeech, and similar general-purpose ASRs, contain the above three listed factors. Our ASR design does not contain the RNN, produces non-sequential outputs (i.e., a single output label per audio sample), and it does not include a large number of neurons/layers. As a consequence, our model architecture consists of a signal processing-based feature extraction layer, followed by multiple convolutional layers. We run experiments with three, four, and five layers containing 100, 150, and 200 hidden units, respectively to observe the impact of model complexity on model accuracy. We also assume a very powerful adversary that knows the~\textit{exact} timestamps of each label in the \captcha{}. This will allow her to pass each label individually to the model, which is known to improve \captcha{} recognition accuracy~\cite{bock2017uncaptcha, dlt}.




\textbf{Training:} We train the model on a batch size of 32, for 50 epochs. We
use early stopping if the cross-entropy loss does not fall more by than 0.01 for
three epochs.
\begin{figure}[!t]
 \includegraphics[width=\linewidth]{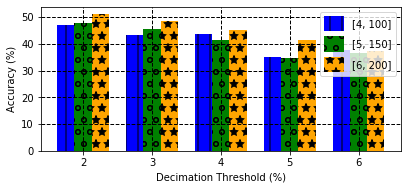}
 \caption{Adversarial training experiments. Increasing $T_d$ decreases the accuracy of the ASR. This is
    understandable since larger $T_d$ results in poorer quality
    audio.~\textbf{LEGEND:} [X,Y]: X = Number of layers. Y = Neurons per layer. \label{fig:adv_training_const_clip}}
\vspace*{-5mm}
\end{figure}

\begin{figure}[!t]                                                                   
 \includegraphics[width=\linewidth]{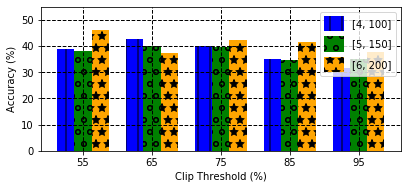}                  
 \caption{Adversarial training experiments. Increasing the $T_c$ decreases the accuracy of the ASR. This is              
    understandable since larger $T_c$ results in poorer quality audio.~\textbf{LEGEND:} [X,Y]: X = Number of layers. Y = Neurons per layer.           
    \label{fig:adv_training_const_dec}}                                          
\vspace*{-8mm}                                                                   
\end{figure}

\textbf{Results:} Figures~\ref{fig:adv_training_const_clip} and
\ref{fig:adv_training_const_dec} show the results of our experiments.
Figure~\ref{fig:adv_training_const_clip} shows the effect of keeping constant $T_c$, while increasing $T_d$. While Figure~\ref{fig:adv_training_const_dec}
shows the impact of increasing $T_c$ while keeping $T_d$ constant. In both
cases, increasing the thresholds will reduce model accuracy.
This is understandable as high thresholds reduce audio quality by discarding
features that are necessary for learning valid decision boundaries.

More importantly, the largest accuracy we get across all experiments is around
51\% for a $T_d$ factor of 2\%~(Figure~\ref{fig:adv_training_const_clip}).  Since audio \captcha{}s are on
average six utterances long, the adversary can
only break the entire audio \captcha{} $0.51^6 = 1.7*10^{-2}$ of the time. This is a
major improvement on the success rate of 94\% observed in prior works~\cite{solanki2017cyber}. A defender can easily
reduce the accuracy rate even further by merely perturbing each utterance with a different $T_d$ and $T_c$ value. The attacker will then be forced to retrain their
model for the highest expected threshold. For example, if
any single utterance in the \captcha{} has a $T_d$ value of
5\%, the attacker will need to train an ASR which has an accuracy
of 41\% in the best case. At this point, the success rate for breaking
\captcha{}s falls further to $0.41^6 = 0.0047$. 


\subsection{Detection}
\label{sec:fingerprinting}

Adversaries often abuse commercial ASR services to transcribe audio \captcha{}s. For instance, Bock et al. showed that an adversary can defeat Google's
reCaptcha~\cite{reCAPTCH28:online} service by feeding its audio \captcha{} to
Google's own ASR~\cite{google_normal}. If ASR owners can detect when an
adversary is misusing their service, they can block the
requests.

\kvus{} is designed to force ASRs to
confuse the \captcha{}s for random noise by outputting only empty strings. Based
purely on the empty transcript, it is not possible for the ASR owner to
ascertain whether an input was real noise or an audio \captcha{}. However, the
key observation here is that random noise and our \captcha{}s have different
levels of acoustic structure. Simply put, random noise lacks any form of
structure, while \kvus{} maintains the structure of the original audio sample.
We hypothesize that due to the difference in these structures, both samples will
produce different sets of activations~\cite{papernot2018deep}.

We analyze the activations produced when passing random noise and \kvus{}
\captcha{}s through an ASR. We start by extracting the activations from each
layer of the ASR for random noise audio samples. These activations correspond to
points in an d-dimensional space, where d is the size of the activation vector.
For each layer, we find the center (i.e., the activation point that has the
smallest $L_2$ distance from all the others). Next, we get the activations for
\captcha{} samples and calculate their distances from the center. If our
hypothesis is correct, noise activations will be closer to the center than the
adversarial one. This will enable us to detect whether an input sample was
adversarial or just noise.



\textbf{Dataset:} We use the same dataset of 56 characters, words, and letters
as before. Our data is split into train and test sets. The train and test set
contained 778 and 600 files respectively. 

\textbf{Models:} This defense method requires white-box access to the ASR since
we are extracting sample activations. The operator of an ASR would realistically
have this level of access to their own model. In our case, we use the DeepSpeech
model, which is a general-purpose ASR and is locally available to us.

\begin{figure}
 \includegraphics[width=\linewidth]{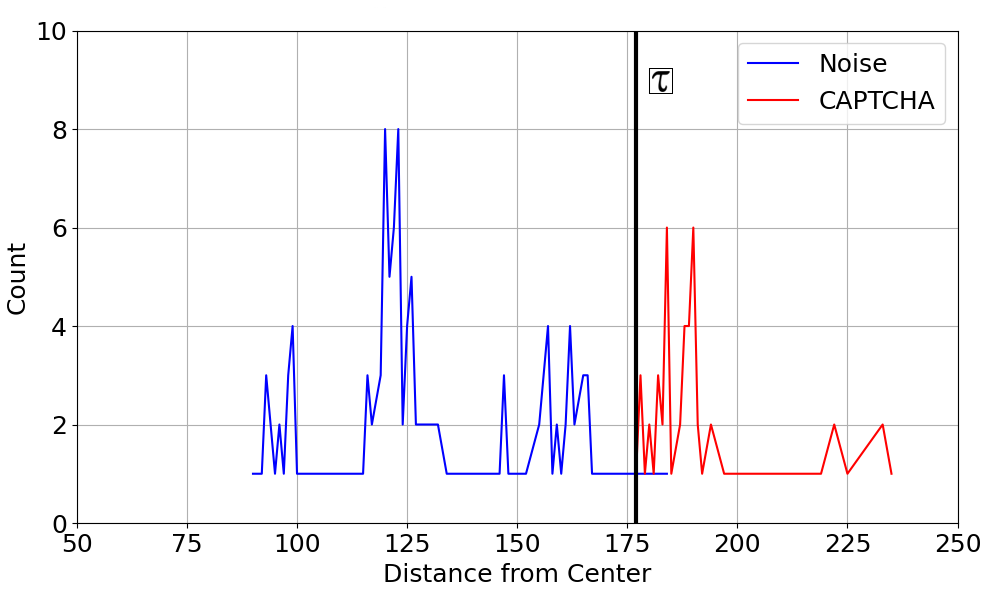}
 \caption{Detection experiment for the
    word ``Forward'' for one layer of the ASR. We can see that \captcha{} have higher distances than those for
    benign examples. Any sample that has an empty output and
    greater distance than the $\tau$ (solid black line), is a \captcha{}.}
 \label{fig:detection}
\vspace*{-8mm}
\end{figure}

\textbf{Results:} Figure~\ref{fig:detection} shows the CDFs for distance values
calculated for both the noise and the \kvus{} sample sets for one layer of the
ASR\footnote{We use all the layers of the ASR as part of the final detection algorithm. However, we are only showing the results for one layer for brevity.}. The figure shows that most of the noise samples (blue) are closer to the
center (which exists at 0,0) than the \captcha{} samples (red). This indicates
that the activations for noise samples are clustered in a small section. In
contrast, \kvus{} activation vectors exist in a larger region. This means we
can use the distance from the center as a simple differentiator between noise
and \captcha{} samples. For example, if the detection threshold, $\tau$, was set at
the value indicated by the black line, activations left of $\tau$ could be
classified as noise, while those on the right could be classified as \captcha{}.
The ASR owner should therefore flag any samples that produce empty string
outputs that have higher activations than $\tau$ as possibly malicious. Since values of $\tau$ are dependent on model activations, the~\textit{exact} value of $\tau$ will depend on the defender's ASR.



However, selecting the optimal $\tau$ is a balancing act and will depend on the
ASR owner's goals. For instance, the aggressiveness of the ASR owner's
mitigation strategy plays a major role in $\tau$'s selection. If the mitigation
strategy is highly aggressive (e.g., account suspension or blocking), then
$\tau$ should be selected to maximize the precision of the detector to ensure
benign users are not being punished.  However, if the mitigation strategy is more
lenient (e.g., delayed response or rate-limiting), then the selected $\tau$
should balance the precision and recall. We select $\tau$ so
that the precision and recall are both equal to 89\%.  Additionally, these
values are calculated for transcribing a single label from our dataset.
\captcha{}s are typically comprised of up to six utterances
(Table~\ref{tab:captcha_service_overview}). An attacker would therefore need to
evade our detector multiple times in order to successfully avoid detection. For
instance, in a six label \captcha{} where a single positive classification is
sufficient to raise suspicious, then the attacker would only have an
$P({\rm evasion}) = (1-0.89)^{6} = 1.77*10^{-4}$\% chance of evading detection. In comparison, the average benign
sample will only have a 18.9\% chance of being incorrectly classified as
suspicious. However, all \kvus{} audio samples only transcribe as an
empty string. Therefore, samples that produce empty strings (an
already rare event for real users) are at risk of being incorrectly labeled
malicious. 

Finally, an attacker could attempt to evade this detection mechanism by further
perturbing the \captcha{} audio. The adversary could perturb the audio randomly
or by employing an optimization technique. In either case, the adversary is
likely to fail since both would adversely affect the transcription of the
underlying audio.

\subsection{Comparison to Existing Literature}

Using the four metrics described in Section~\ref{sec:kv2_eval}, we now compare our work against existing methods from current literature. The idea of using adversarial machine learning algorithms to build robust \captcha{} is still new. Very little work has been done in this space. Therefore, we were only able to identify two papers~\cite{shekhar2019exploring, choi2018poster} that attempt to build robust \captcha{} algorithms using adversarial machine learning algorithms. While these works present good first steps in this space, they have significant limitations:

\textbf{Transferability:} Both these methods have very low transferability rates against black-box ASRs. This allows the attacker to easily reconstruct the CAPTCHA text, specifically, with rates of 36\% and 48\% (for \cite{shekhar2019exploring} and \cite{choi2018poster} respectively) compared to our method which is $<$ 0.01\%.

\textbf{Audio Quality:} It is not possible to assess their audio quality since the authors perform very small user studies. \cite{shekhar2019exploring} ran the study with 15 participants, and \cite{choi2018poster} ran it with 10 participants. In contrast, our user study included around 200 participants across different age groups and backgrounds. 

\textbf{Detection and Adaptive Adversary:} Neither paper provides experiments regarding detection or adaptive adversaries. Our audio has distinctive characteristics that can help the ASR owner detect if their service is being abused. Additionally, we show that our method is robust to adaptive adversaries.

\section{Related Work}
\label{sec:relwork}
Audio \captcha{}s are an essential
tool for protecting online resources. However, they can be easily broken using ASRs \cite{bock2017uncaptcha}. To prevent this, audio CAPTCHAs must be
created to be robust to ASRs, while simultaneously ensuring human intelligibility. 
One way to do so would be via adversarial samples. These are imperceptibly modified inputs that can force ML models to produce a misclassification
\cite{goodfellow2014explaining,szegedy2013intriguing, kurakin2016adversarial,baluja2017adversarial,brown2017adversarial,sharif2016accessorize,papernot2016limitations,nguyen2015deep,zhao2017generating,song2018constructing}.
One of the most popular ways of producing these samples against ASRs is via use of optimization attacks\cite{schonherr2018adversarial,cai2018attacking,gong2017crafting,kereliuk2015deep,yuan2018commandersong,alzantot2018did,kreuk2018fooling,cisse2017houdini,kumar2018skill,carlini2018audio,yakura2018robust,qin2019imperceptible}. However, these attacks require white-box access to the target ASR, which limits their use for \captcha{} generation
purposes. 

To add to that, these attacks fail to demonstrate transferability~\cite{sok} (i.e., surrogate and target ASRs produce different outputs). This alone does not rule out optimization attacks as viable CAPTCHA algorithms. This is because the attack audio might produce one of two transcriptions: phonetically similar or dissimilar. For CAPTCHA audio, we want a phonetically dissimilar output (e.g., ``HadJSNm'' for the audio containing ``one two three''). As a result, the bot's ASR will be unable to reconstruct the original CAPTCHA text. However, if the attack audio output is phonetically similar to the original CAPTCHA command (e.g., ``Juan too tree'' for the audio containing ``one two three''), then the bot can easily reconstruct the \captcha. Prior work~\cite{sok} failed to identify which one of the two was the case. In this paper, we show that optimization attacks produce phonetically similar transcriptions, making them vulnerable to Bots.

Due to these aspects, the \captcha{} algorithms~\cite{shekhar2019exploring, choi2018poster} that build on these optimization attacks suffer from severe limitations, which include lack of effectiveness against black-box models, questionable audio quality, and robustness against adaptive adversaries. To overcome these, we propose a \captcha{} generation that exploits the differences in how the human ear and ASRs process audio. As a result, our attack can successfully exploit~\textit{any} ASR, produce high-quality audio that is easy for human listeners to understand, while also being robust to adaptive adversaries.

Additionally, we demonstrate how adversarial example attacks against ASR models can be
used to create secure and usable audio CAPTCHAs. Using an adversarial example
attack to generate \captcha{}s has been previously proposed for text and image
\captcha{}s \cite{shi2019adversarial,kwon2020robust,osadchy2017no}. These
\captcha{}s were shown to be nearly impossible to transcribe by machine, but
still usable by humans. 

\section{Conclusion}
\label{sec:conc}
Breaking audio \captcha{}s has become easier with the improvement in machine
learning systems. Specifically, adversaries can now use ASRs to correctly
transcribe \captcha{}s. In response, \captcha{} services have been forced to
degrade the quality of their audio, which has adversely impacted human
intelligibility. We design a new attack, \kvus{}, that can produce intelligible
audio \captcha{}s that are also robust to ASRs. We evaluate it against a range
of criteria, including intelligibility, robustness to adaptive adversaries, and
transferability. We show that no attack in this space
fulfills all of these requirements and that our method is the only viable audio
\captcha{} generator currently in existence.
\vspace{10mm}
\small
\bibliographystyle{plain}
\bibliography{main}
\appendix
\section{Appendix}
\label{sec:app}
\begin{figure}
 \includegraphics[width=\linewidth]{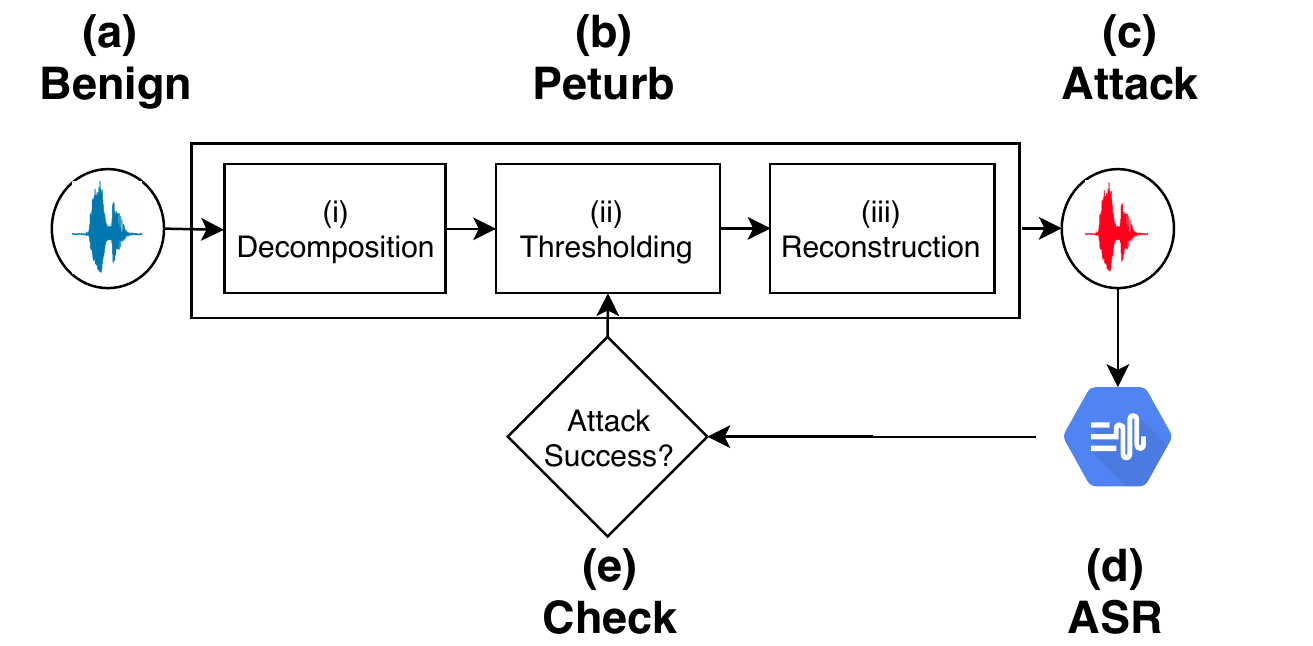}
 \caption{The \kvog{} attack pipeline. (a) We start with the audio sample we
    wanted to perturb. (b(i)) It is decomposed into frequency components using
    the DFT. (b(ii)) Next, all the frequency bins that have less power than
    the threshold are set to zero. (b(iii)) The remaining frequencies are
    reconstructed.  (c) This produces a raw perturbed audio, (d) which is passed
    to the ASR. (e) If the attack succeeds, the threshold value is decreased,
    otherwise, the threshold is increased. Then the steps repeated again.}
 \label{fig:baseline}
\vspace*{-5mm}
\end{figure}

\begin{table}[]

\begin{tabular}{|c|c|c|c|}
\hline
\textbf{Service} & \textbf{\begin{tabular}[c]{@{}c@{}}Potential\\ Captcha\\ Use\end{tabular}} & \textbf{Min Len} & \textbf{Max Len} \\ \hline
Apple \cite{appledeveloper} & Numbers (10-99) & 3 & 5 \\ \hline
BotDetect \cite{BotDetec57:online} & 0-9 a-z & 4 & 6 \\ \hline
Captchas.net \cite{captchas34:online} & 0-9 alpha-zulu & 4 & 6 \\ \hline
Microsoft \cite{chellapilla2005designing} & Sound Identifcation & 3 & 3 \\ \hline
Securimage \cite{Securimage} & 0-9 a-z & 4 & 6 \\ \hline
Telerik \cite{telerik.com} & 0-9 alpha-zulu & 4 & 6 \\ \hline
\begin{tabular}[c]{@{}c@{}}Google\\ (Recaptcha v1) \cite{von2008recaptcha}\end{tabular} & 0-9 & 10 & 10 \\ \hline
\begin{tabular}[c]{@{}c@{}}Google\\ (Recaptcha v2) \cite{von2008recaptcha}\end{tabular} & words & 3 & 3 \\ \hline
\begin{tabular}[c]{@{}c@{}}Google\\ (Recaptcha v3 ) \cite{von2008recaptcha}\end{tabular} & words & 3 & 3 \\ \hline
Ebay  & Recaptcha & Recaptcha & Recaptcha \\ \hline
MTCaptcha \cite{mtcaptcha} & 0-9 a-z & 4 & 4 \\ \hline
\multirow{2}{*}{Amazon } & Recaptcha & Recaptcha & Recaptcha \\ \cline{2-4} 
 &Sound Identifcation & 3 & 3 \\ \hline
Slashdot & Recaptcha & Recaptcha & Recaptcha \\ \hline
\end{tabular}

\caption{An overview of the existing commercial \captcha{} services. Since
    words, numbers, digits are commonly used by these services, we use them in
    our experimental setup. We also base our results on \captcha{}s of length 6,
    since this is the most common size. \label{tab:captcha_service_overview}}

\end{table}

\end{document}